\documentclass[preprint,showpacs,onecolumn,preprintnumbers,amsmath,amsfonts,amssymb,floatfix,aps,superscriptaddress]{revtex4}

\usepackage{graphicx}
\usepackage{epsfig}
\usepackage[hang,nooneline]{subfigure}
\usepackage[normalem]{ulem}
\usepackage{color}
\usepackage{hyperref}

\newcounter{fig}

\begin{document}

\title{Vector Dark-Antidark Solitary Waves in
Multi-Component Bose-Einstein condensates}
\author{I. Danaila}
\email[Email: ]{ionut.danaila@univ-rouen.fr}
\affiliation{Laboratoire de Math{\'e}matiques Rapha{\"e}l Salem, Universit{\'e}
de Rouen, 76801 Saint-{\'E}tienne-du-Rouvray,  France}
\author{M.~A.~Khamehchi}
\affiliation{Washington State University,
Department of Physics \& Astronomy,
Pullman, WA 99164 USA}
\author{V.~Gokhroo}
\affiliation{Washington State University,
Department of Physics \& Astronomy,
Pullman, WA 99164 USA}
\author{P.~Engels}
\affiliation{Washington State University,
Department of Physics \& Astronomy,
Pullman, WA 99164 USA}
\author{P. G. Kevrekidis}
\affiliation{Department of Mathematics and Statistics, University of Massachusetts
Amherst, Amherst, MA 01003-4515, USA}

\date{\today} 

\begin{abstract}
Multi-component Bose-Einstein condensates exhibit an intriguing variety of nonlinear structures.
In recent theoretical work, the notion of magnetic solitons has been introduced.
Here we generalize this concept to vector dark-antidark solitary waves
in multi-component Bose-Einstein condensates. We first provide
concrete experimental evidence for such states in an atomic BEC and subsequently illustrate the broader concept of these states, which are based on
the interplay between miscibility and inter-component repulsion. Armed with this more general conceptual
framework, we expand the notion of such states to higher dimensions
presenting the possibility of both vortex-antidark states and
ring-antidark-ring (dark soliton) states.
We perform numerical continuation studies, investigate the existence of these states and examine their stability using the method of Bogolyubov-de Gennes analysis.
Dark-antidark and vortex-antidark states are found to be 
stable for broad parametric regimes. In the case
of ring dark solitons, where the single-component ring state is known
to be unstable, the vector entity appears to bear a progressively more
and more stabilizing role as the inter-component coupling is increased.
\end{abstract}

\maketitle

\section{Introduction}

Atomic Bose-Einstein condensates (BECs) offer an excellent
testbed for the exploration of waveforms
relevant to multi-component nonlinear wave systems~\cite{stringari,siambook}.
A principal paradigm consists of the dark-bright (DB) solitary wave and related structures such as dark-dark solitary waves  that have long been studied theoretically~\cite{christo,vdbysk1,vddyuri,ralak,dbysk2,shepkiv,parkshin}.
The experimental study of such states was pioneered much earlier
in nonlinear optics, including e.g. the observation of dark-bright
solitary wave structures in~\cite{seg1,seg2}. Yet, it was the
versatility and tunability of BECs that enabled a wide variety
of relevant studies initially motivated by the proposal of~\cite{buschanglin}.
Specifically, the experimental realization of DBs~\cite{hamburg}
was followed by a string of experiments investigating the dynamics and properties of these features
including in-trap oscillations of DBs, their spontaneous generation (e.g.
via counterflow experiments) and their interactions both with other
DBs and with external potential barriers~\cite{pe1,pe2,pe3,pe4,pe5,azu}.

Very recently, a different type of multi-component solitons was proposed,
the so-called ``magnetic solitons'' \cite{stringa}. Due to the limited number of solitonic families that have been proposed so far, and due to the even fewer number of types observed in experiments, such entities naturally are of great theoretical interest. The ability to generate them using current state-of-the art experiments with multi-component BECs gives them considerable experimental appeal as well.
These states have a complementary intensity profile in the two-components
$(\psi_1(x,t), \psi_2(x,t))$
and are described by the two-component wave function
\[ \left( \begin{array}{ccc}
\psi_1(x,t)\\
\psi_2(x,t) \end{array} \right)
= \sqrt{n} \left( \begin{array}{ccc}
 \cos(\frac{\theta}{2}) e^{i \phi_1}\\
\sin(\frac{\theta}{2}) e^{i \phi_2} \end{array} \right)
\]
where $\theta(x,t)$ characterizes the spatial distribution
of the amplitude, $n$ the total density and $\phi$ the phase of each 
component. It is relevant to note 
that a related idea regarding the ansatz of the multi-component nonlinear wave
state was put forth, e.g., in the work of~\cite{fetter}.

A complementary possibility recognized considerably earlier was that of
dark-antidark solitary waves~\cite{epjd}. Antidark solitary waves
are bright solitary waves on top of a finite background.
Here, we will avoid calling the structures under investigation ``magnetic'',
as we do not a priori constrain the sum of the densities of the two
components to equal that of a single component ground state, as in the
settings of~\cite{stringa,fetter}. We will show that the
idea of complementary non-trivial components, one of which is
antidark, is a very general one and is applicable in several dimensions as well:
similar ideas naturally emerge in two dimensions in the form of vortex-antidark and ring-antidark-ring states, which to the best of our understanding have previously not been explored.
We motivate and 
complement our theoretical prediction, numerical verification and stability
analysis of such states with an example of an experimental realization in a BEC
that confirms that dark-antidark states are straightforward to
create and observe the dynamics of in current experimental settings.

The fundamental rationale behind such states is
reminiscent of that of the DB entities:
For a single-component system, an extensive discussion of the existence and stability of excited states such as dark solitons or vortices can be found in the respective $1d$ and $2d$ chapters of~\cite{siambook}.
In a two-component system with inter-component repulsion, a dark soliton or a vortex in one component will induce a potential in the second component. If now atoms of the second (``bright'') component are added {\it in the absence} of a spatially extended background of the second component, the density suppression in the first component will get filled by atoms of the second component and
 a dark-bright, a vortex-bright~\cite{kody,pola}
or a ring-DB solitary wave~\cite{ringjan} will emerge.
However, if the second component features a (spatially
extended) ground state profile, the presence of inter-component
repulsion will produce an effective additional potential which will attract atoms of the second component into the dip of the first
one. This generates a bright solitary wave {\it on top of} the
existing nontrivial background, forming an antidark solitary wave.
An additional constraint in this case is that the two components need to coexist in the ``wings''
(i.e., sides) of the dark-antidark (DA), vortex-antidark (VA),
or ring-antidark-ring (RAR) structure.
This imposes the condition of miscibility between
the two species~\cite{timm,Pubig,aochui}, i.e. the condition that the 
 inter-component repulsion should be less than the square root of the product
of the intra-component ones, $0 \leq g_{12} < \sqrt{g_{11} g_{22}}$.
We note in passing here that this condition is derived in the
context of homogeneous BECs and is only slightly affected by the
presence of weak trapping conditions as in the case examples 
considered herein~\cite{rafael}.

Based on the discussion above, there is a straightforward path
that one can follow in order to establish such states
(at least, numerically) involving an antidark
component. One can start at the uncoupled limit of $g_{12}=0$
with an excited state (e.g., a dark soliton in $1d$, or a vortex
or a ring dark soliton in $2d$) in one component and a fundamental (ground)
state in the second component. Then, after turning on the inter-component
coupling, the dip of the excited state in the first component will induce an effective
attracting potential (due to the inter-component repulsion) in the second component, attracting some atoms into the dip while maintaining (due to 
miscibility) the background of the second component. By construction, an antidark structure is formed.

Such a state, as we will see in detail below, will continue
to exist for values of $g_{12}$ up to  the miscibility-immiscibility
threshold. To discuss these types of states, we will proceed as follows:
 in section II, we will provide an example for an experimental realization
of a dark-antidark solitary wave that will serve as a key motivation for the
corresponding theoretical more in-depth study. In section III,
we will explore the relevant states numerically, using numerical
continuation and bifurcation theory, starting from the uncoupled limit
described above. Finally, in section IV, we will summarize our
conclusions and present some intriguing possibilities for future work.

\section{Experimental Results}

\begin{figure}[tb]
\begin{center}
\includegraphics[width=\textwidth]{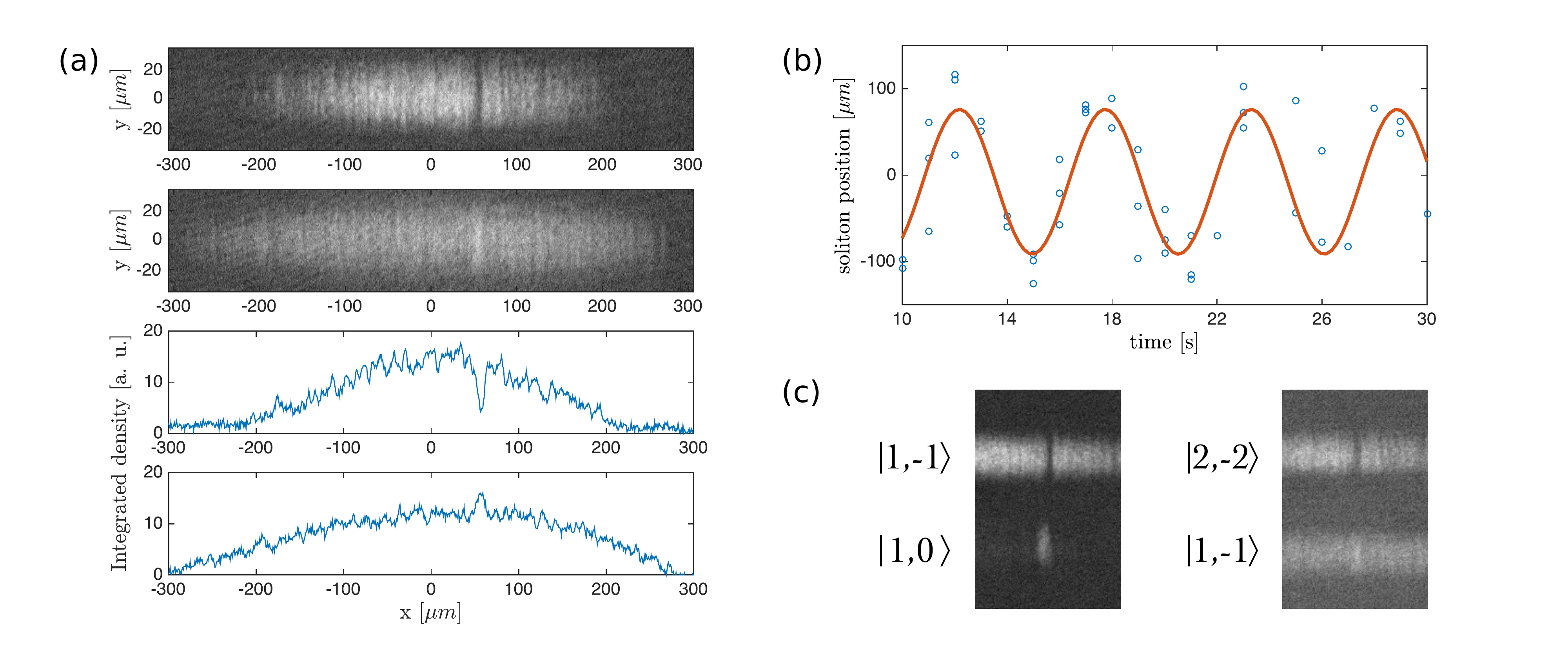}
\caption{(Color online) Experimental realization of dark-antidark solitary waves. (a) Absorption images (upper two panels) and corresponding integrated cross sections (lower two panels) of a dark-antidark solitary wave. The dark soliton component resides in a cloud of $ |F,m_F\rangle=|2,-2 \rangle$ atoms (upper and third panel), while the bright component consists of atoms in the $ |F,m_F\rangle=|1,-1 \rangle$ state (second and forth panel from top). (b) Experimentally observed oscillation of the dark-antidark solitary wave in the trap. The position is measured along the x-axis, i.e. along the weakly confining axis of the trap. The time is measured starting from the initial microwave pulse that creates the two-component mixture. The blue dots are experimental data, while the red line is a sine function fit to the data. (c) Comparison between a dark-bright soliton in a mixture of atoms in the $ |F,m_F\rangle=|1,-1 \rangle$ and $ |F,m_F\rangle=|1,0 \rangle$ states (left image) and a dark-antidark structure in a mixture of atoms in the $ |F,m_F\rangle=|2,-2 \rangle$ and $ |F,m_F\rangle=|1,-1 \rangle$ states (right image).
}
\label{dad_fig1}
\end{center}
\end{figure}

To motivate our discussion, we begin by presenting experimental evidence for the existence,
stability and dynamics of a dark-antidark solitary wave, shown in Fig.~\ref{dad_fig1}.
In our experiments we observe these features in two-component
BECs confined in an elongated dipole trap. The experiments begin by
creating a BEC of approximately $0.8\times 10^6$ $^{87}$Rb atoms held in an
optical trap with harmonic trap frequencies of
$\omega_{x,y,z} = 2\pi \{1.4, 176, 174\}$~Hz, where z is the direction of gravity. Evaporation in the dipole trap is
continued until no thermal fraction is discernible. Initially, all atoms
are in the $|F,m_F \rangle = |1,-1 \rangle$ hyperfine state.
Subsequently, a brief microwave pulse transfers a fraction of the atoms
(approximately 50\% for the case shown in Fig.~\ref{dad_fig1}) into the
$|2,-2 \rangle$ hyperfine state. The transfer occurs uniformly across the
whole BEC. The resulting two-component BEC can be described by two separate
Gross-Pitaevskii equations that are only coupled by the inter-component
scattering length (see the theory section below). The intra- and
inter-component scattering lengths for the two states are $a_{11}=100.4 a_0$,
$a_{22} = 98.98 a_0$, and $a_{12} = 98.98 a_0$, where $a_{11}$
denotes the scattering of two atoms in the $|1,-1 \rangle$ state, $a_{12}$
the scattering between an atom in the $|1,-1 \rangle$ and $|2,-2 \rangle$
state, and $a_{22}$ between two atoms in the $|2,-2 \rangle$
state \cite{kokk}; $a_0$ is the Bohr radius. Based on the standard miscibility
argument~\cite{timm,Pubig,aochui},
discussed above, this mixture is slightly miscible. 
However, the difference between $a_{11}$ and $a_{22}$ leads to a slight concentration of $|2,-2\rangle $ atoms towards the center of the cloud.
When the mixture is held in trap for approximately 10~sec or longer, the emergence
of a dark-antidark solitary wave is observed as shown in Fig.~\ref{dad_fig1}. The solitary waves are
imaged by suddenly switching off the trap and imaging the $|2,-2 \rangle$ state after
10~ms of expansion and the $|1,-1 \rangle$ state after 11~ms of expansion. The difference
in the free-fall time separates the two images on the camera, so that the two
components appear below each other in the images. During all in-trap evolution leading
to the soliton formation, the two components have been well overlapped vertically.
Repeating this procedure with well controlled experimental parameters, we observe that each
iteration of the experimental run reliably produces a two-component BEC containing one
dark-antidark solitary wave such as the one shown in Fig.~\ref{dad_fig1}.  In all iterations,
the dark soliton resides in the $|2,-2 \rangle$ component and the antidark soliton is found in the $|1,-1
\rangle$ component.

It is remarkable that these features emerge quite ``naturally'' in our experiments without any
dedicated wavefunction engineering~\cite{pe1,pe4} or phase imprinting~\cite{hamburg}. 
Furthermore, these features are very long lived. We have observed their in-trap dynamics for up to 30 sec.
For comparison, starting with a $50/50$ mixture of the two components, we measure the lifetime of the $|2,-2 \rangle$ component to be
$\sim 22$~sec and that of $|1,-1 \rangle$ component to be $\sim 33$~sec for our experimental parameters.
The emergence of these solitary waves is rather insensitive to the exact mixture ratio of
the components. Experimentally we tested and confirmed their existence in a variety of mixtures ranging from $30$\% of the atoms in the $|2,-2 \rangle$ state and $70$\% in the $|1,-1 \rangle$, to mixtures of $50$\% in the $|2,-2 \rangle$ state and $50$\% in the $|1,-1 \rangle$ state. In mixtures where the abundance of the $|2,-2 \rangle$
component exceeded approximately $50$\%, no clear soliton formation was observed.

We have also repeated the experiment with a mixture of atoms in the $|1,-1 \rangle$ and $|1,0 \rangle$ components. The scattering lengths for this mixture are $a_{11} = 100.4 a_0$,
$a_{22} = 100.86 a_0$, and $a_{12} = 100.41 a_0$, where $a_{11}$ now
denotes the scattering of two atoms in the $|1,-1 \rangle$ state, $a_{12}$
the scattering between an atom in the $|1,-1 \rangle$ and $|1,0 \rangle$
state, and $a_{22}$ between two atoms in the $|1,0 \rangle$
state \cite{kokk}. This mixture is closer to the miscibility-immiscibility threshold. Following an analogous procedure as described above, no formation of dark-antidark solitary waves was observed. Instead, dark-bright solitons were generated. A comparison between a dark-bright soliton and a dark-antidark one shown in  Fig.~\ref{dad_fig1}(c), showcasing their very different structure in the bright component. This emphasizes the important role that the miscibility of the component plays for the generation of a non-zero background in the second (bright) component, as has also been highlighted in~\cite{stringa}.

The long lifetimes and reproducible generation of dark-antidark solitary waves in a mixture of atoms in the $|2,-2 \rangle$ and $|1,-1 \rangle$ states allow us to observe their in-trap dynamics (Fig.~\ref{dad_fig1}(b)). We clearly detect a slow oscillation of the solitary wave along the weak axis of the trap. A fit of the data in Fig.~\ref{dad_fig1}(b) gives an oscillation period of approximately 5.6~sec. For comparison, the period of a dark soliton in a single-component BEC in the same trap is predicted to be 1~sec \cite{busch2000,konotop}. Hence, the dark-antidark solitary waves are significantly slower. A similar trend has been found in ~\cite{buschanglin,pe2}, where it was seen that dark-bright solitons are slowed down when the amount of atoms in the bright component is increased. The theory of ``magnetic solitons'' described in \cite{stringa} assumes $a_{11} = a_{22}$, which in our experiment is only approximately fulfilled. This theory predicts an oscillation period on the order of 9.8~sec, somewhat longer than that observed in the experiment. A quantitative comparison between experiment and theory, including the influence of the mixing ratio of the two components and the finite lifetime of the trapped atoms, will be left for future work. Here, these first observations of dark-antidark solitary waves serve as a motivation to investigate (chiefly as a function of the inter-component scattering length) 
and generalize the underlying concepts of dark-antidark structures using 
numerical continuation studies and Bogolyubov-de Gennes analysis.

\section{Theoretical/Numerical Results}

In order to capture the qualitative features of the states of interest, it will suffice
to utilize a mean-field model in the form of the Gross-Pitaevskii equation. Upon suitable
standard reductions~\cite{stringari,siambook}, the model can be transformed 
to its dimensionless version in the form:
\begin{eqnarray}
i \frac{\partial \psi_1}{\partial t} &=& -\frac{1}{2} \Delta \psi_1 + V({\bf x}) \psi_1 +
\left(g_{11} |\psi_1|^2 + g_{12} |\psi_2|^2 \right) \psi_1
\label{eqn1}
\\
i \frac{\partial \psi_2}{\partial t} &=& -\frac{1}{2} \Delta \psi_2 + V({\bf x}) \psi_2 +
\left(g_{12} |\psi_1|^2 + g_{22} |\psi_2|^2 \right) \psi_2
\label{eqn2}
\end{eqnarray}
Here, the pseudo-spinor field is denoted by $(\psi_1,\psi_2)^T$ (where $^T$ is used for transpose),
$V({\bf x})= \frac{\Omega^2}{2} {\bf x}^2$ represents the parabolic trap, while $g_{ij}$
are the interaction coefficients, proportional to the  experimental scattering lengths
mentioned above. In line with the analysis of~\cite{stringa}, we will assume in what
follows that $g_{11}=g_{22}=g$, while $0 \leq g_{12} \leq g$. Given that only the ratios
of the scattering lengths matter, we will choose $g=1$, while $0 \leq g_{12} < 1$, in
order to be in the miscible regime, while preserving inter-component
repulsion.

In our numerics, the stationary states $(\psi_1^{(0)},\psi_2^{(0)})^T$ are identified by virtue
of a fixed point iteration (typically a Newton method) originally at $g_{12}=0$,
i.e., the limit where the two components are uncoupled.
Then, parametric continuation is utilized in order to follow the configuration
as a function of $g_{12}$ up to the miscibility threshold.

Upon computing the solution, Bogolyubov-de Gennes stability analysis is implemented
that perturbs the solutions according to:
\begin{eqnarray}
	\psi_{1} &=& e^{-i \mu_1 t} \left( \psi_1^{(0)}({\bf x}) + \delta (a({\bf x}) e^{i \omega t}
	+ b^{\star}({\bf x}) e^{-i \omega^{\star} t}) \right)
	\label{deq3}
	\\
	\psi_{2} &=& e^{-i \mu_2 t} \left( \psi_2^{(0)}({\bf x}) + \delta (c({\bf x}) e^{i \omega t}
	+ d^{\star}({\bf x}) e^{-i \omega^{\star} t}) \right)
	\label{deq4}
\end{eqnarray}
Here $\omega$ represents the linearization eigenfrequency and
the vector $(a,b,c,d)^T$ is the linearized eigenvector pertaining to the respective
eigenfrequency. The chemical potentials are denoted by 
$(\mu_1,\mu_2)$, while $\delta$ is a formal small parameter of the linearization
ansatz. $\Omega$ represents
the strength of the trapping potential in the longitudinal vs. the transverse directions (i.e., the ratio thereof) and thus
needs to be $\Omega \ll 1$ for the reductions to be meaningful. In the 
following we will assume
$\Omega=0.2$ and $\mu_1=\mu_2=2$ unless noted otherwise.

{Numerical computations are performed  using FreeFem++ \cite{hecht-2012-JNM}. The numerical system developed for computing stationary solutions of the Gross-Pitaevskii equation \cite{dan-2010-JCP} was extended for the two-component  system (\ref{eqn1})-(\ref{eqn2}).
	We use quadratic ($P^2$) finite elements with mesh adaptivity, offering high-resolution of the steep gradients in the solution.  The Bogolyubov-de Gennes linear eigenvalue problem  corresponding to the $P^2$ finite element discretization is solved using  the ARPACK  library.}

\subsection{1d: Dark-Antidark Solitary Waves}

\begin{figure}[tb]
\begin{center}
\includegraphics[scale=0.5]{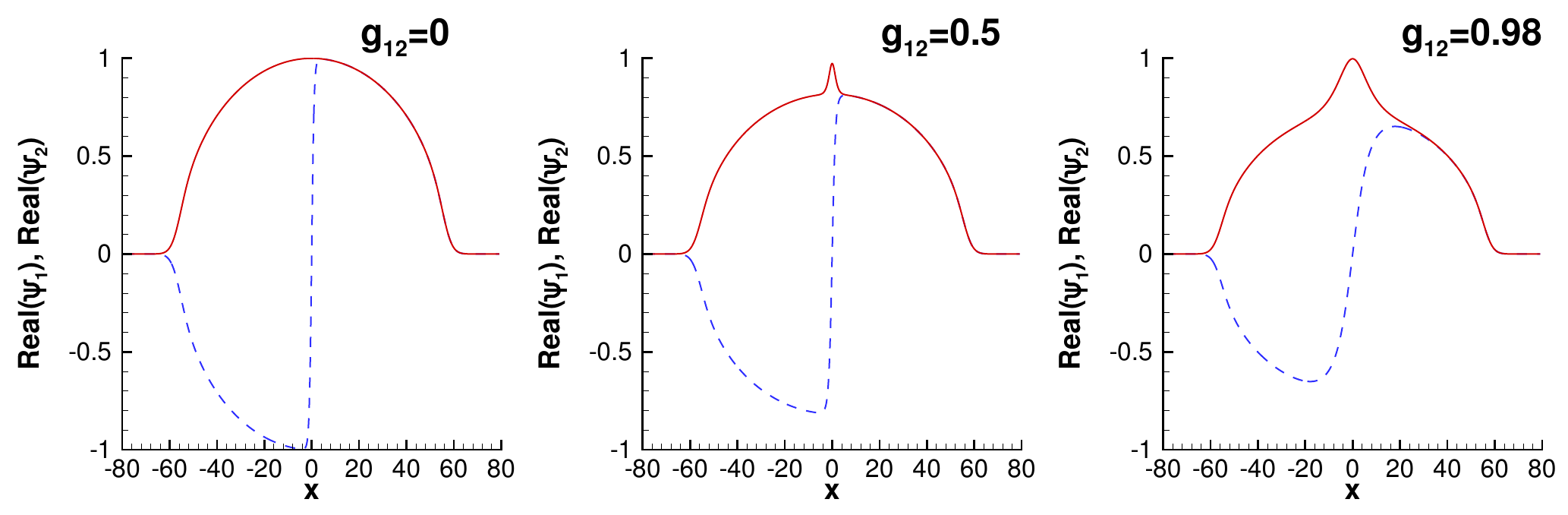}
\includegraphics[scale=0.5]{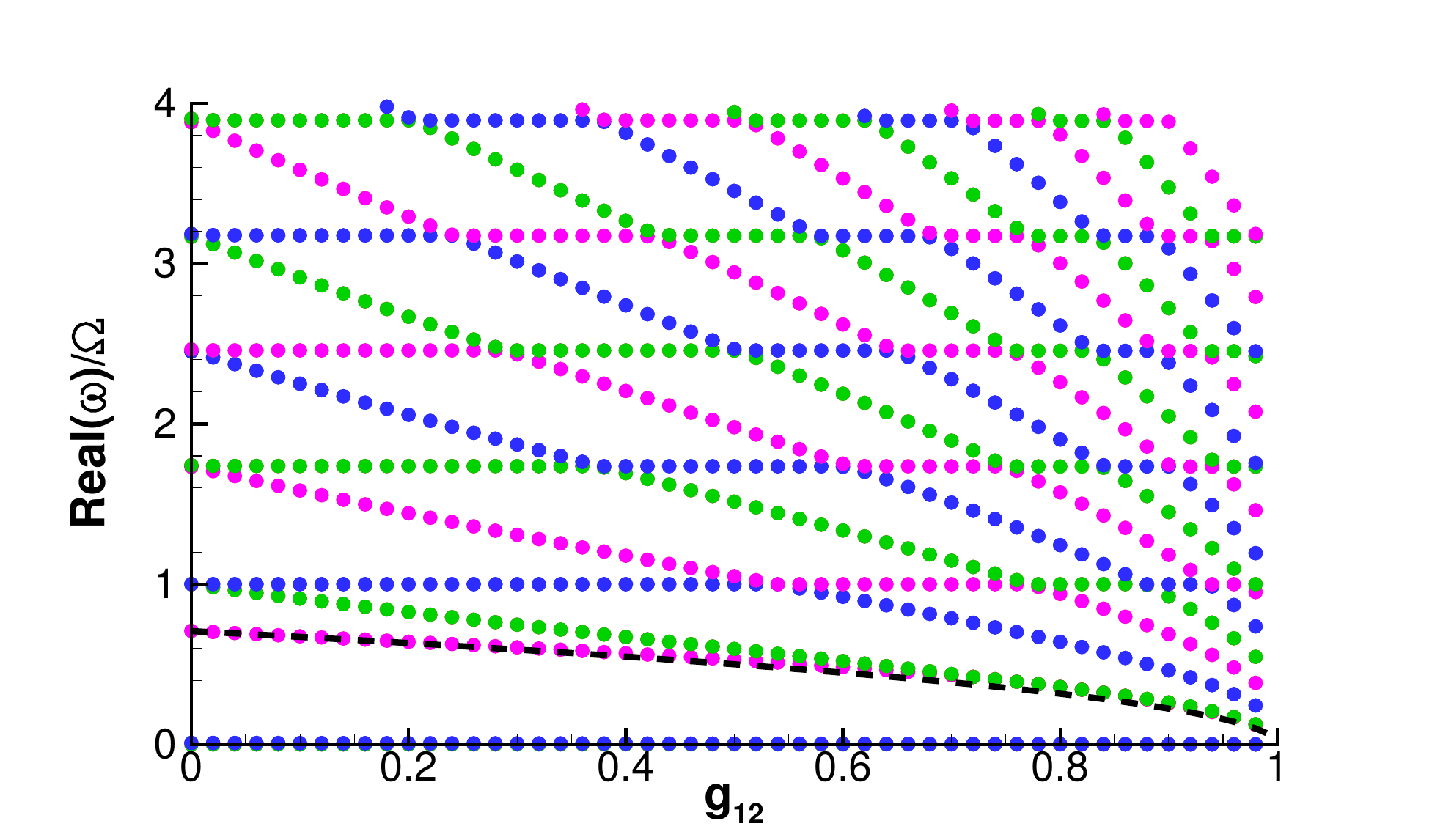}
\caption{(Color online) The top panel of the figure 
shows 3 examples of the two components in
the dark-antidark state for progressively increasing
$g_{12}$, i.e., for $g_{12}=0$, $g_{12}=0.5$, and
for $g_{12}=0.98$ 
The bottom panel shows the dependence of the lowest
eigenfrequencies scaled by the trap frequency $\Omega$ which in this 1d example
is chosen to be $\Omega=0.025 \ll 1$. The dashed black line indicates the theoretical
prediction for the anomalous mode (see discussion in the text). 
The colors in the bottom panel are there only to visually aid the eye
to identify the continuation of the different modes.}
\label{dad_fig2}
\end{center}
\end{figure}

We start by considering the scenario of dark-antidark solitons in one spatial
dimension. Solutions of this type are represented in Fig.~\ref{dad_fig2}.

For $g_{12}=0$ the solution has the form of a regular dark
solitary wave coupled to a fundamental state in the second component. For a finite value of
$g_{12}$ the stationary solution develops a bump at the location of
the dark soliton dip. Due to the inter-component repulsion, the density dip in the the first component
leads to an attracting potential well for the second component. Therefore an antidark peak is formed that becomes stronger as $g_{12}$ is increased, while
the first component tends to vanish as the miscibility-immiscibility threshold
of $g_{12}=1$ is approached. This trend is clearly seen in the top panel of
Fig.~\ref{dad_fig2}. For this case, we have used $\Omega=0.025$, although
similar results have been found for other values of the trap strength.

Remarkably, in the case of the dark-antidark solitary wave family we find the relevant solution
to be generally stable (as shown in the right panel of Fig.~\ref{dad_fig2}) through
a wide interval of variation of the $g_{12}$ parameter; an
extremely weak oscillatory instability arises for $0.71 < g_{12} < 0.87$,
that will be discussed further below, yet its growth rate is so small
that we do not expect it to affect the dynamics in an observable
way over the time scales of interest. In the large chemical
potential limit, we in fact have a detailed handle on the spectrum of the relevant
eigenfrequencies in an analytical form. When $g_{12}=0$, the second component is uncoupled from the first and its spectrum in the ground state  consists of eigenfrequencies
\begin{eqnarray}
\omega= \sqrt{\frac{n (n+1)}{2}} \Omega
\label{eqn5}
\end{eqnarray}
where $n$ is a non-negative integer, as discussed in~\cite{stringa3,pelirecent}.
The dark soliton (DS) spectrum on the other hand, as explained, e.g., in~\cite{siambook} consists
of the spectrum of the ground state in which the soliton is ``embedded'', given by
Eq.~(\ref{eqn5}), as well as an extensively studied 
anomalous (or negative energy) mode
associated structurally with the excited state nature of the DS state, and practically
with its oscillation inside the parabolic trap. As is well known, the latter mode has the frequency
$\omega=\Omega/\sqrt{2}$ in the large chemical potential 
limit~\cite{busch2000,konotop} 
and is the lowest excitation frequency in the system.
Using an argument similar to that presented in~\cite{stringa}, we expect that the
relevant mode scales as $\omega \approx \Omega \sqrt{\delta g/(2 g)}$, where $\delta g=g - g_{12}$.
This prediction is also represented in the right panel of Fig.~\ref{dad_fig2} and is in reasonable
agreement with the corresponding numerical result throughout the interval of variation
of $g_{12}$. Regarding the rest of the spectrum (the modes associated with the ground
state in each component), it can be seen that they can be partitioned into two fundamental
categories, namely those that are essentially left invariant and those that undergo a rapid monotonic decrease (which is nearly linear for small $g_{12}$) as $g_{12}$ increases. As this takes place, it is in principle possible
for the anomalous mode (of dark soliton oscillation) and the lowest
frequency associated with the ground state to collide and lead to a
resonant eigenfrequency quartet~\cite{siambook}. This does happen
in the example of Fig.~\ref{dad_fig2} for $0.71 < g_{12} < 0.87$,
yet as mentioned above the growth rate of this instability is miniscule
and hence it will not be considered further herein.

\subsection{2d: Vortex-Antidark Solitary Waves}

\begin{figure}[tb]
	\begin{minipage}{0.4\textwidth}
		\begin{center}
	\includegraphics[width=\textwidth]{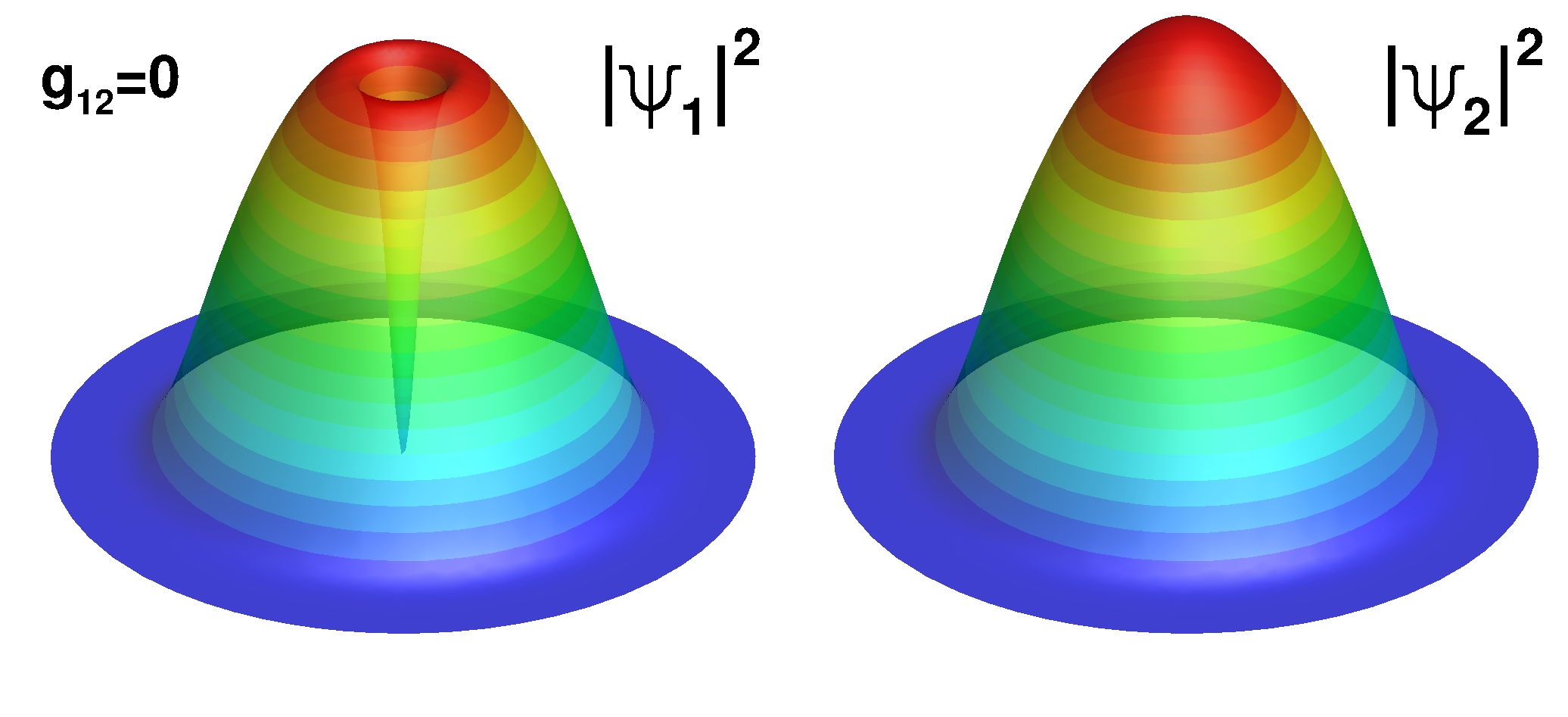}
	\includegraphics[width=\textwidth]{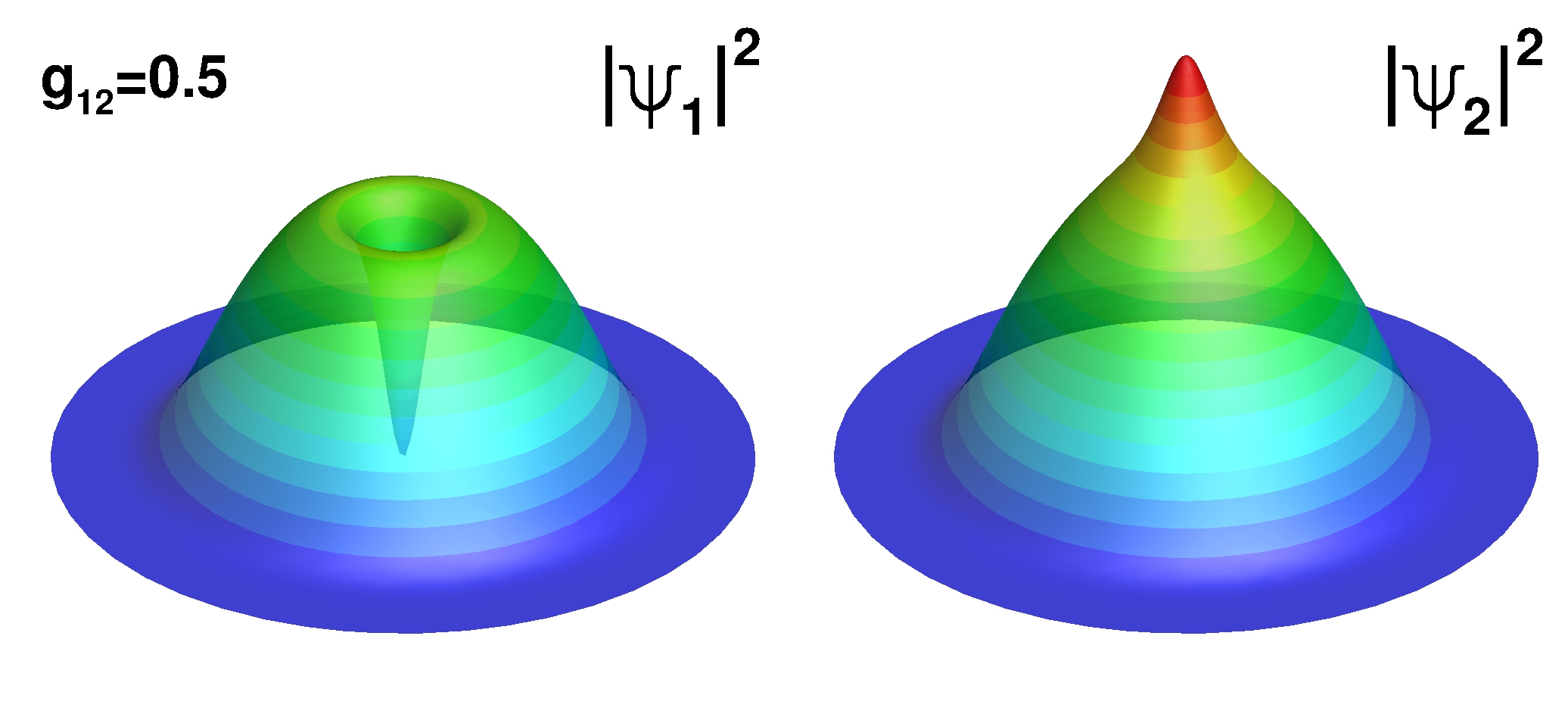}
	\includegraphics[width=\textwidth]{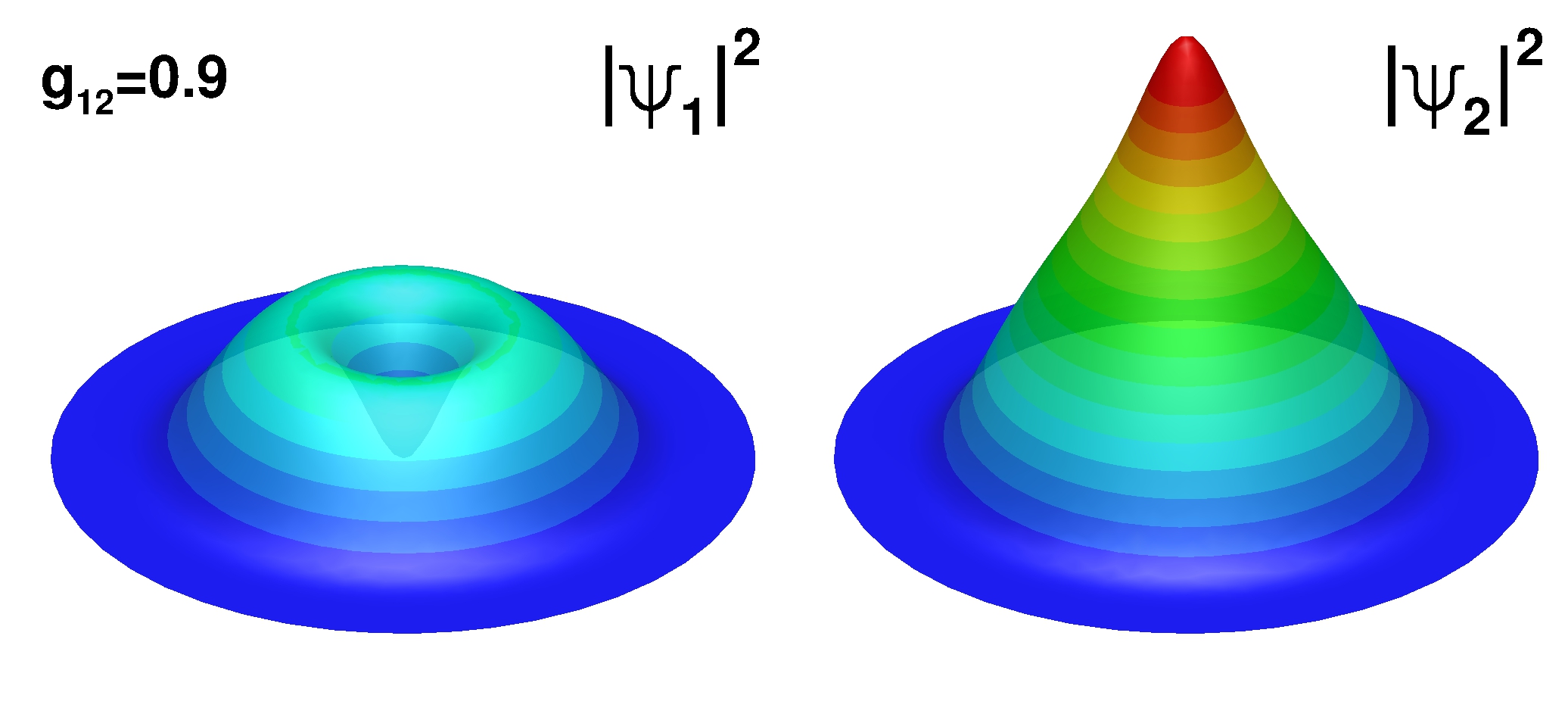}
\end{center}
	\end{minipage}\hfill
	\begin{minipage}{0.5\textwidth}	
\begin{center}
\includegraphics[width=\textwidth]{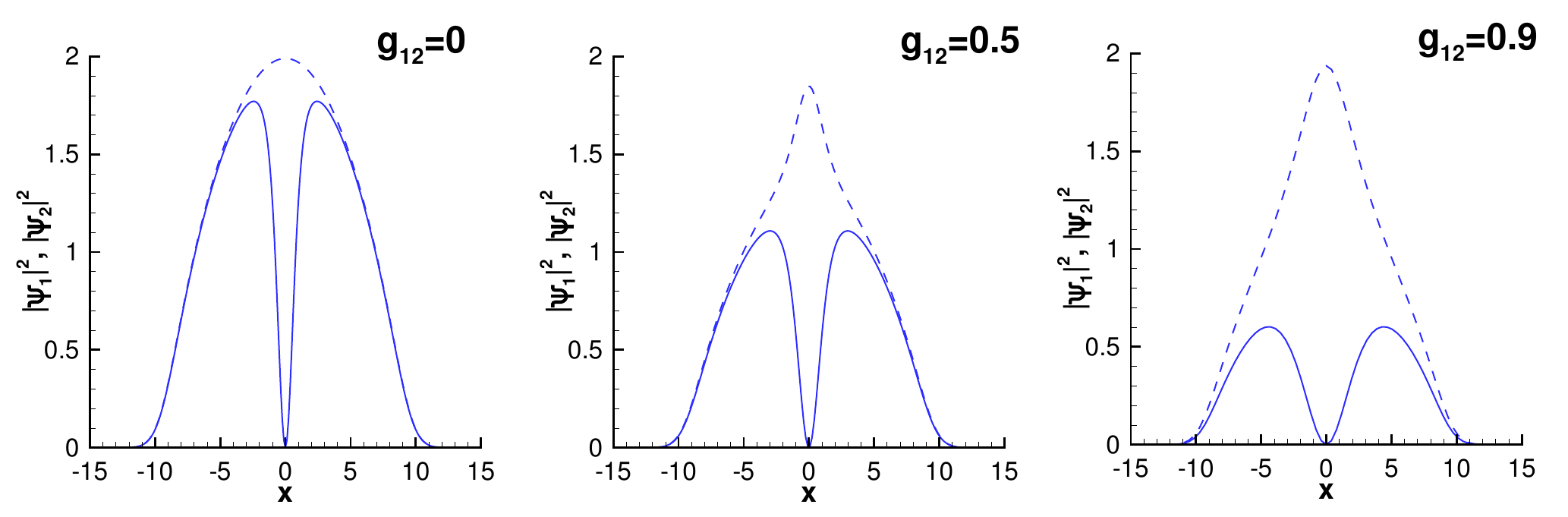}
\includegraphics[width=\textwidth]{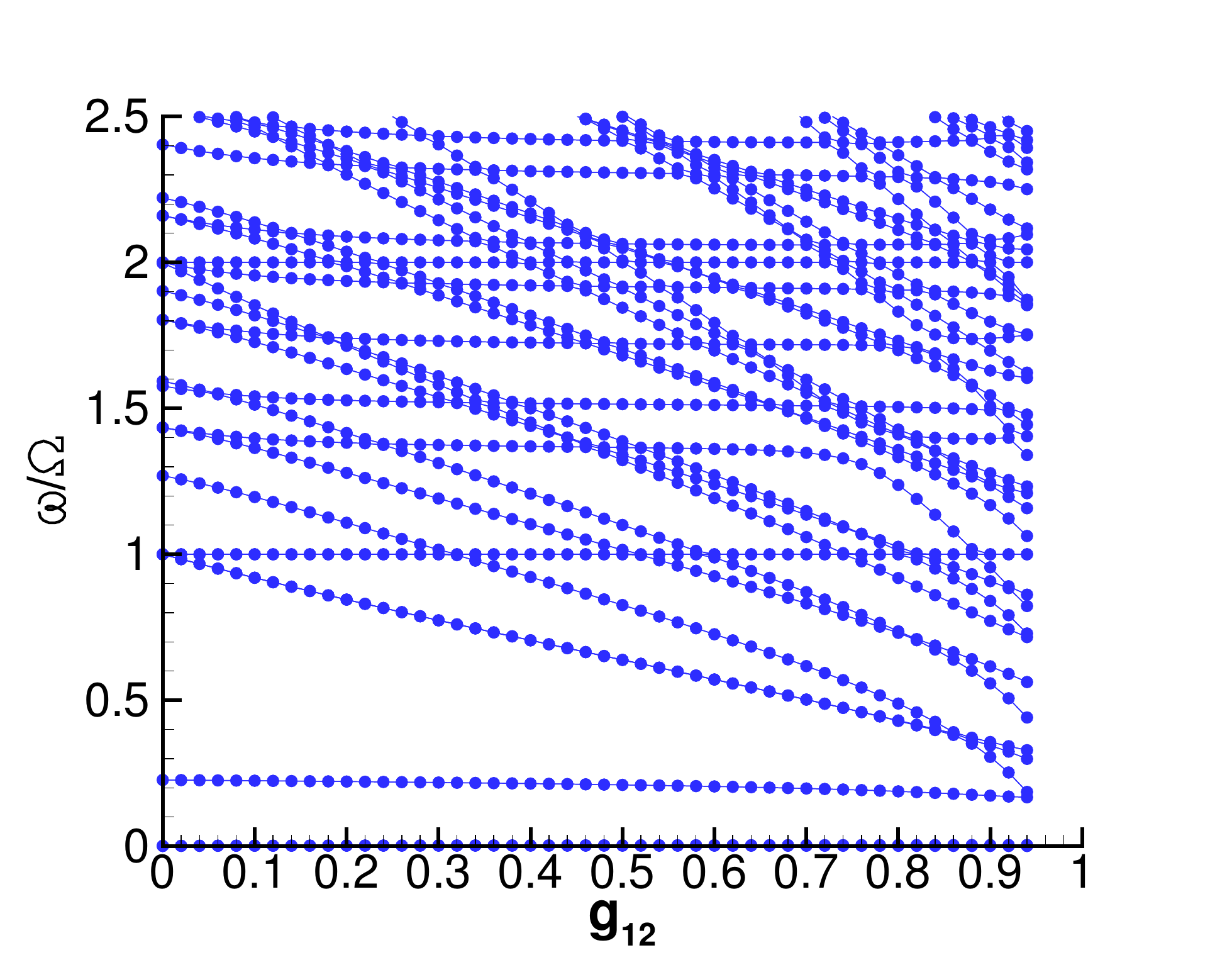}
\end{center}
	\end{minipage}
		
\caption{(Color online) In the left panels of the 
figure, we see three-dimensional renderings
of the density as a function of $x-y$ for the vortex-antidark states and
for the particular values of $g_{12}=0$, $0.5$ and $0.9$. 
The top right panel illustrates for completeness a cut through 
the density of these states (at $y=0$).
Lastly, the bottom
right panel illustrates how the frequencies of the spectral BdG analysis
``evolve'' as $g_{12}$ is varied (see also the relevant detailed discussion in the text).
Notice that all frequencies remain real over the parametric interval
considered, indicating the spectral stability/dynamical robustness
of the associated state.}
\label{dad_fig3}

\end{figure}

We now generalize the above concept to the 2d case, as is shown in Fig.~\ref{dad_fig3}.
Our first example in this setting replaces the 1d dark soliton by a 2d vortex in the
first component that generates an attractive potential well for the second component.
Once again, it can be seen that starting from the decoupled limit of $g_{12}=0$ and
increasing $g_{12}$ in the miscible regime, the coupled vortex-antidark state emerges
with progressively more and more atoms of the second component being radially concentrated
in the well formed by the vortex. This creates a radially symmetric
antidark solitary wave in the second component.

For this case, the spectrum is shown in the bottom right panel of the figure and also features
two sets of constituents. In the limit of $g_{12}=0$, the ground state component
consists of frequencies theoretically
approximated (in the Thomas-Fermi limit of large $\mu$ considered
here) by
\begin{eqnarray}
\omega= \sqrt{m + 2 k  (1+m) + 2 k^2} \Omega,
\label{eqn6}
\end{eqnarray}
where $k, m \geq 0$ are non-negative integers. The first component bearing the vortex
also shares these frequencies (due to the vortex being again ``embedded'' within the
ground state), but additionally carries an anomalous (negative energy) mode that is a signature
of its excited state nature~\cite{siambook}. This mode is associated with the rotation of
the vortex around the center of the parabolic trap. It is well characterized
by the frequency $\omega=\frac{\Omega^2}{2 \mu} \log(A \mu/\Omega)$, where
$A \approx 2 \sqrt{2} \pi$~\cite{stephan}. As $g_{12}$ is increased, a similar trend as in 1d is observed: one set of frequencies remains
invariant, while a second set, originally degenerate with the first at $g_{12}$=0, monotonically decreases
as $\delta g \rightarrow 0$. The anomalous mode frequency associated with the vortex monotonically decreases as well, although in a less pronounced
manner.

Physically, this implies that the additional atoms stored within the antidark
component decrease the rotation frequency (i.e., increase the rotation period) of the composite
entity within the trap. A similar behavior was found for dark-bright solitons
even experimentally~\cite{buschanglin,pe2}. Thus, for dark-antidark solitons and for vortex-antidark
solitary waves, ``heavier means slower''.

\subsection{2d: Ring-Antidark-Ring Solitary Waves}

Finally, to illustrate the generality of the underlying concept, we explore the
role of increasing the inter-component (repulsive) coupling $g_{12}$ in the miscible
regime for the case where the first component bears a ring dark soliton (RDS), while the
second one is started in the ground state, again at large $\mu$ (i.e., in 
the vicinity of
the Thomas-Fermi limit). Ring dark solitons have been predicted early on in the context of BECs~\cite{george},
following their proposal and even experimental observation in nonlinear optics~\cite{kivsharyang,dreischuh}.
Subsequent detailed studies of their stability~\cite{carr,cluster} illustrated that they are unstable
for all values of the chemical potential from the linear limit onwards, progressively becoming
more unstable as $\mu$ is increased. The initial instability is towards a quadrupolar mode
leading to $4$ vortices, while subsequently hexapolar (leading to 6 vortices), 
octapolar (leading to 8 vortices)
etc. instabilities arise in the relevant spectrum.

\begin{figure}[tb]
	\begin{minipage}{0.4\textwidth}
		\begin{center}
			\includegraphics[width=\textwidth]{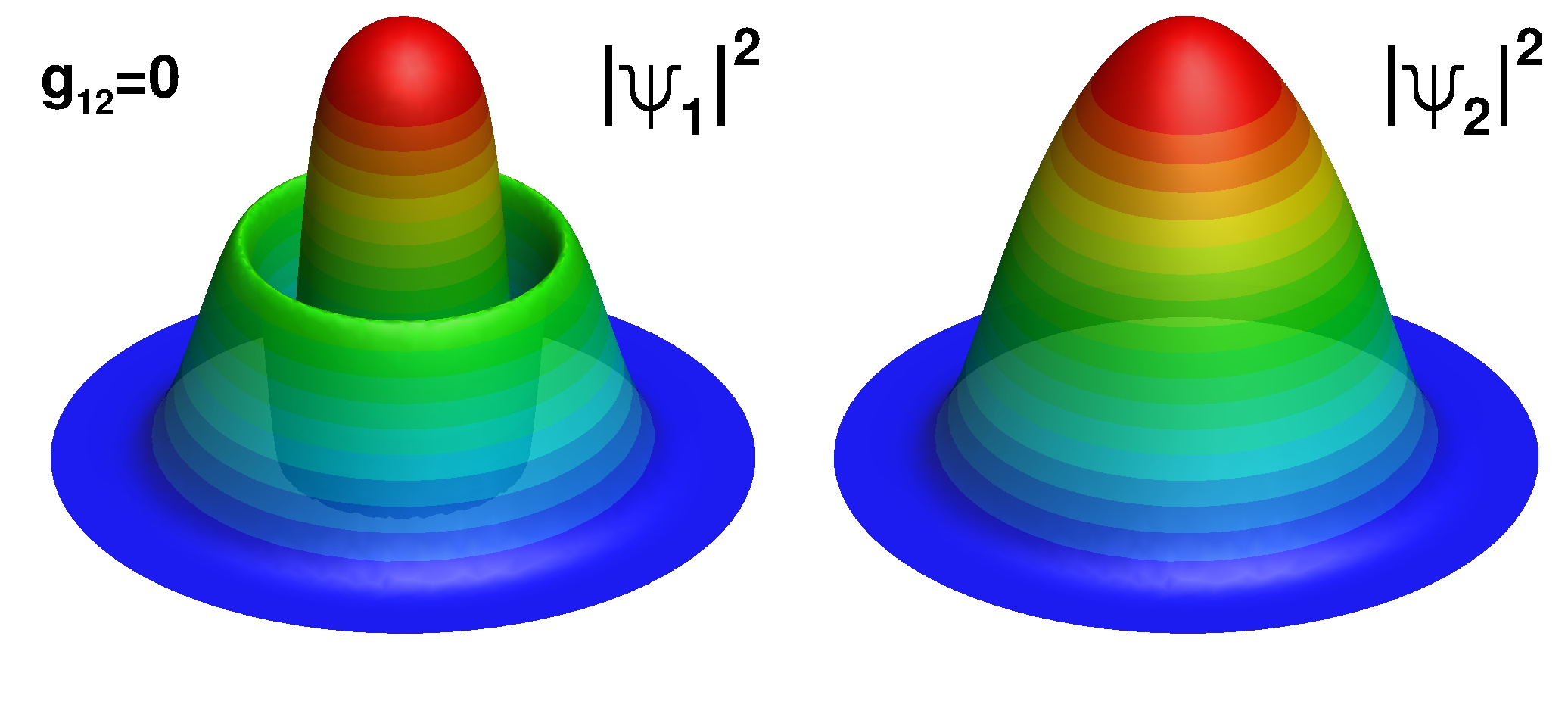}
			\includegraphics[width=\textwidth]{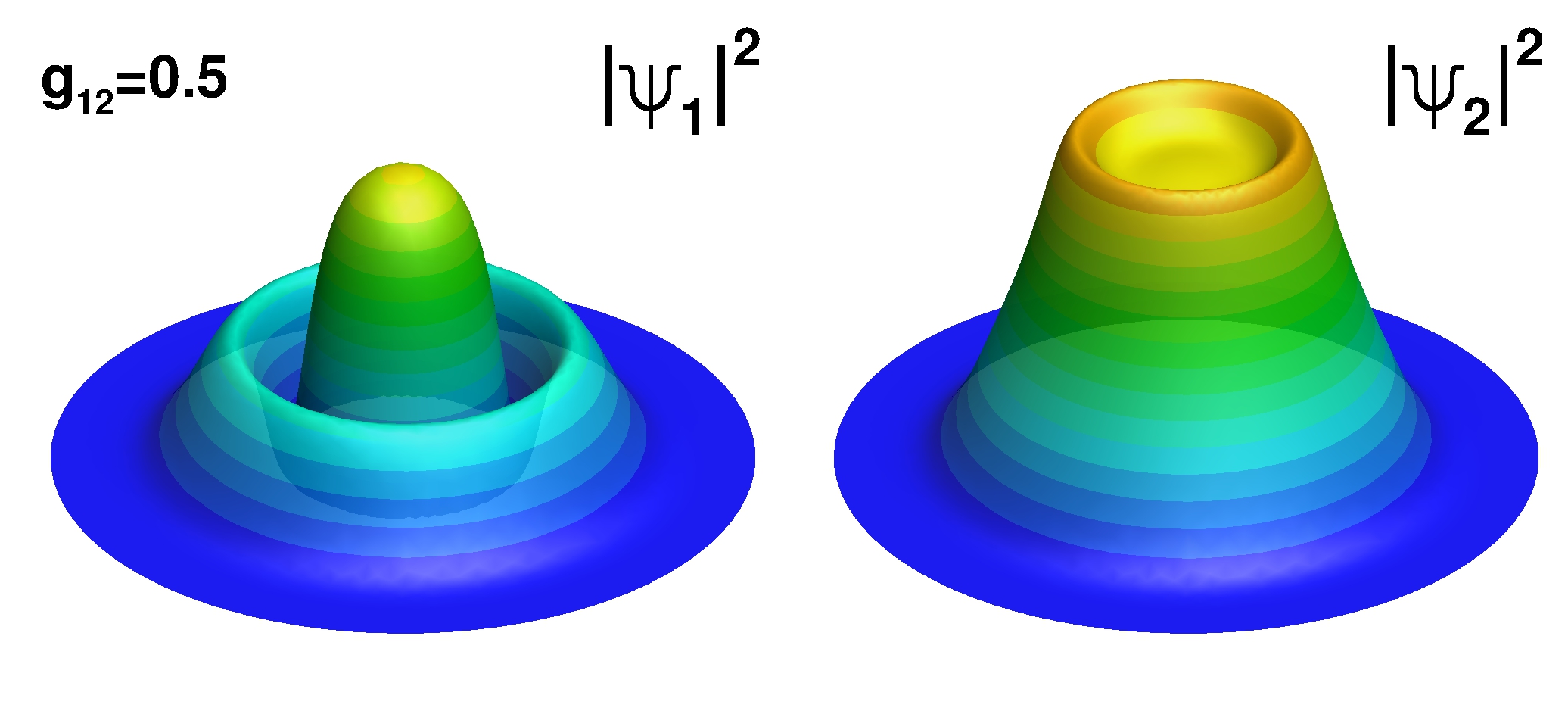}
			\includegraphics[width=\textwidth]{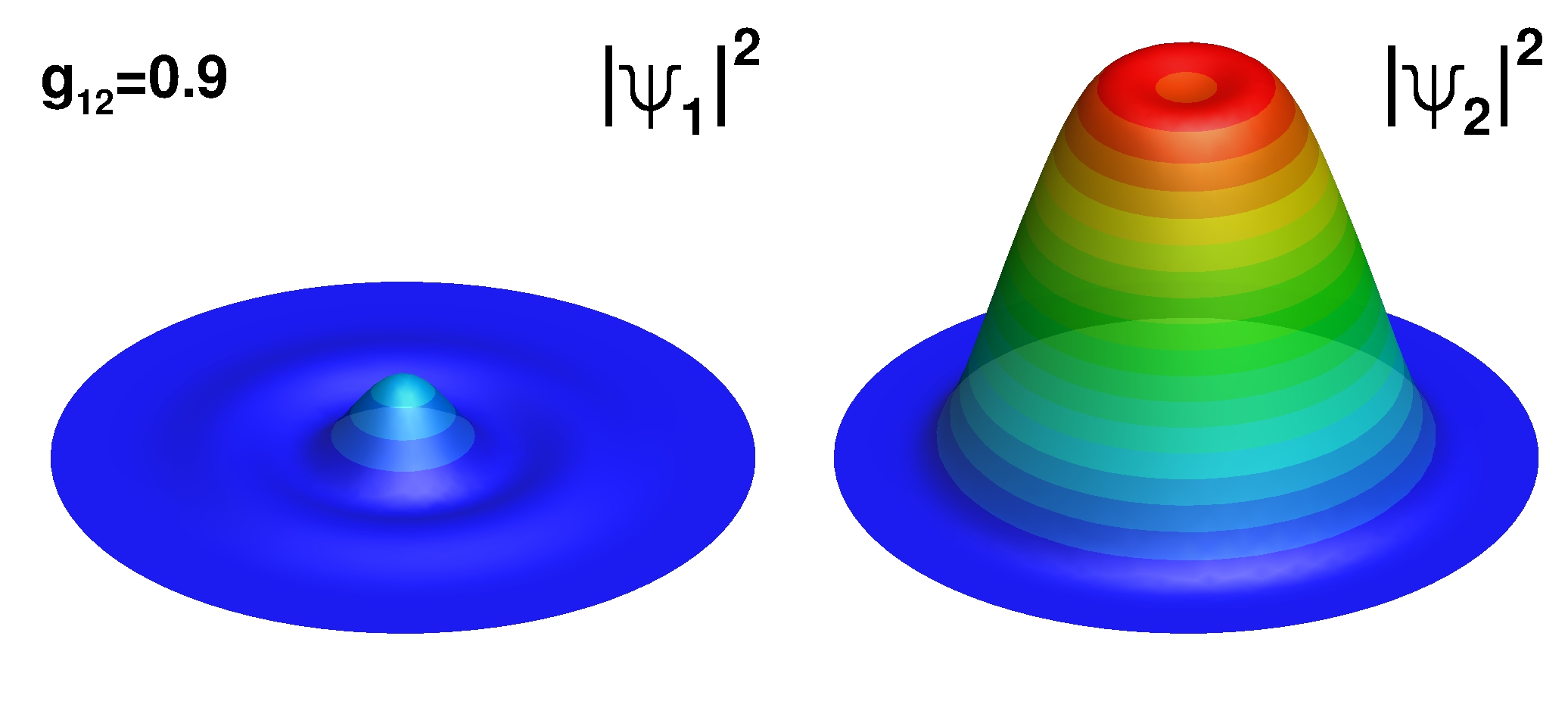}
		\end{center}
	\end{minipage}\hfill
	\begin{minipage}{0.5\textwidth}	
		\begin{center}
			\includegraphics[width=\textwidth]{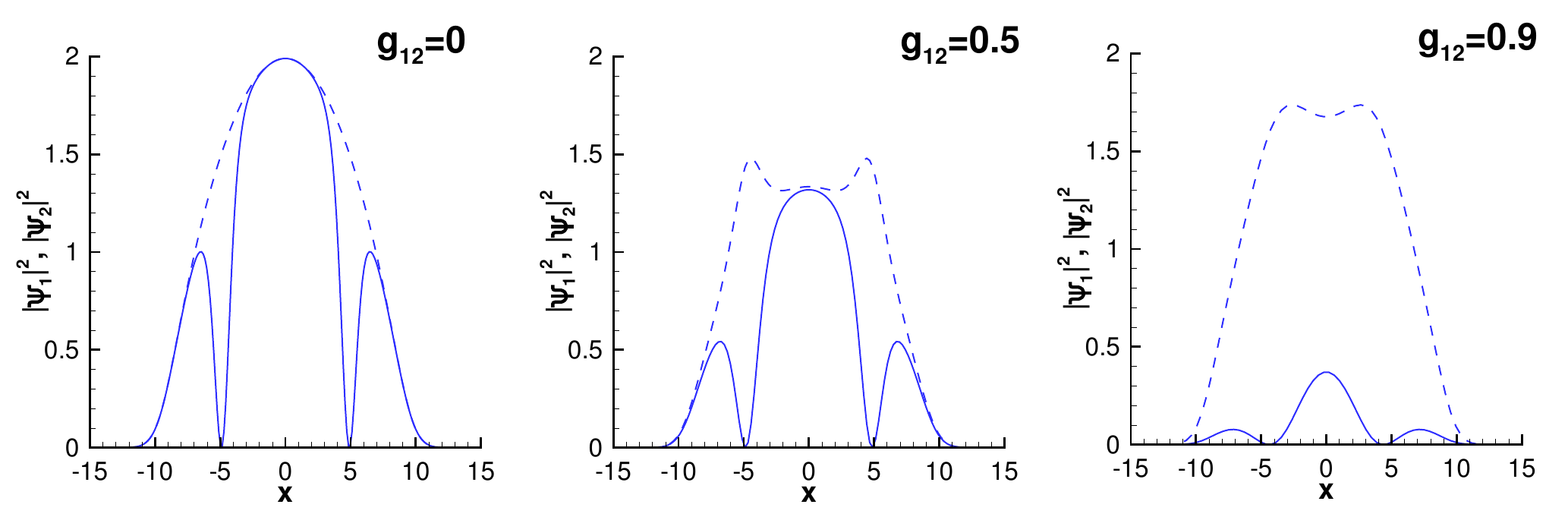}
			\includegraphics[width=\textwidth]{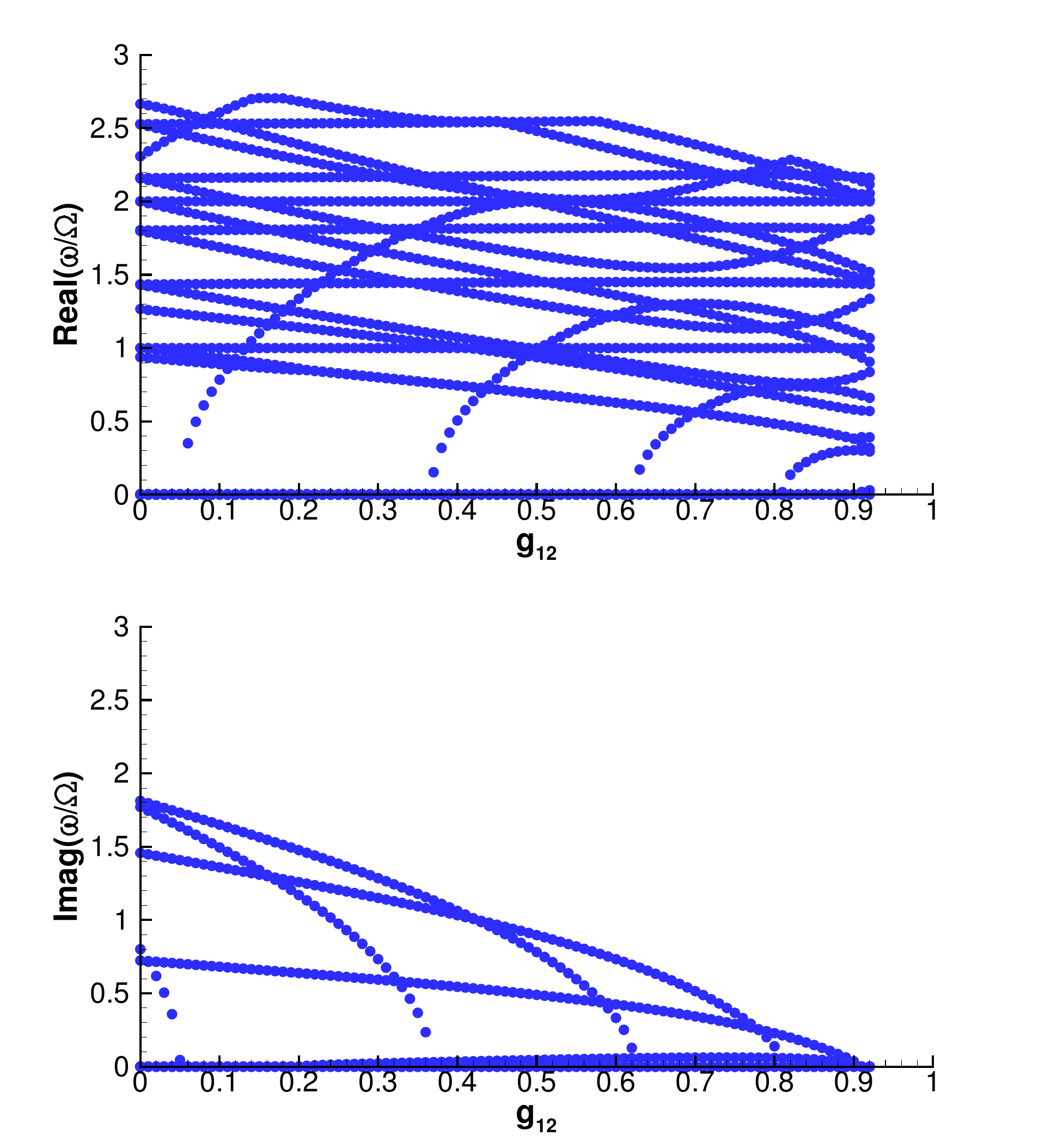}
		\end{center}
	\end{minipage}	

\caption{(Color online) Same as the previous figure, but now
for the dark-antidark ring state. The left panel presents
the 3d profiles of the state and the top right illustrates the cross
sections (for $y=0$). The bottom right panel shows the real and
imaginary parts of the linearization eigenfrequencies of the system
normalized by the trap frequency of
$\Omega=0.2$. Notice
how the imaginary eigenfrequencies associated with instabilities
to (azimuthal) snaking progressively disappear as $g_{12}$ is increased.
I.e., the ring antidark solitary wave has a stabilizing effect on the
unstable ring dark soliton.}
\label{dad_fig4}
\end{figure}

Here, we observe the relevant states in Fig.~\ref{dad_fig4}.
A progressive increase of $g_{12}$ generates an attractive annular potential for atoms in the second component
so that an antidark ring is formed in the second component.
As with the previous states, the increase of $g_{12}$ also decreases
the density of the dark ring component (for the same chemical potential), finally leading it towards
extinction as the miscibility-immiscibility threshold is approached. An associated and quite interesting
feature is that as $g_{12}$ is increased, progressively more and more unstable
modes of the original single component RDS are {\it stabilized}, hence it is intriguing to note
that for the same chemical potential states with larger $g_{12}$ and hence larger antidark component
are less susceptible to instability. In fact, the state tends towards complete stabilization as
the miscibility threshold is approached, a feature inter-related with the tendency towards extinction of
the associated RDS component. It is relevant to note also that in these
last two cases (of Figs.~\ref{dad_fig3} and~\ref{dad_fig4})
the state appears to terminate {\it noticeably before}
the homogeneous miscibility limit of $g_{12}=1$, while the 1d corresponding
state can be identified continuously up to the miscibility-immiscibility
threshold.

\section{Conclusions \& Future Work}

In the present paper, we propose a multitude of states motivated by the recent proposal of magnetic solitons
put forth in ~\cite{stringa}. These ``dark-antidark'' states that we discuss are
based on a simple physical principle, namely the formation of an attractive potential
well on top of a ground state component by the presence of a ``dark'' structure in the
other component. In the miscible case with inter-component repulsion, 
this potential well attracts atoms and forms an antidark entity
(a 1d soliton, a 2d soliton or even a ring soliton) in the (formerly) ground state component.
This is a natural generalization also of the notion of dark-bright solitons which have
recently been extensively explored. 

We have proposed this notion
at an intuitive/theoretical level and have illustrated its generality via detailed numerical
computations. Furthermore, we have showcased it in experiments, at least in as far as its 1d installment
is concerned. This emphasizes the relevance of these features for current experiments with multi-component BECs. 
An additional appealing feature of such symbiotic structures lies in the
fact that when the single component entity (such as the ring dark soliton) may be unstable,
this coupling appears to have a stabilizing effect rendering the relevant entity more amenable
to observation. Both the existence and the spectral stability characteristics of these
states were explained over variations of the inter-component coupling throughout their
range of existence.

Naturally, there are many open directions for future study in this subject. From an
experimental perspective, it would be particularly interesting to explore the possibility
to form such states in both two- and three-dimensions. In the latter setting of 3d,
computations would also be especially useful in elucidating such states: recently,
vortex-line-bright and vortex-ring-bright~\cite{stathis} states have been 
identified,
and their generalization to antidark ones would be quite relevant, as well as the
study of their stability.
From a theoretical perspective, it would also be quite intriguing to explore further the
``particle description'' of such entities, both at the level of the single particle
(e.g. characterizing the rotation of a vortex-antidark solitary wave etc.), 
but also quite importantly
at the level of interaction of multiple such entities. The latter has not been quantified
even in 1d settings. As a final but important point, we note the sensitive
dependence of the emergence in experiments of the dark-antidark
vs. dark-bright solitons on the value of scattering lengths in the 
vicinity of the miscibility threshold. This is a feature worth
further elucidating in future experiments and corresponding theoretical
analyses.
Some of these topics are presently under consideration and will
be reported in future publications.

\acknowledgments P.G.K. gratefully acknowledges the support of NSF-DMS-
1312856, and the ERC
under FP7, Marie Curie Actions, People, International
Research Staff Exchange Scheme (IRSES-605096). He also acknowledges
the warm hospitality of the Laboratoire de Math{\'e}matiques Rapha{\"e}l Salem
at the University of Rouen. P.E. gratefully acknowledges funding from NSF under grant number PHY-1306662.
{ID  acknowledges support  from the French ANR grant  ANR-12-MONU-0007-01 BECASIM (Mod{\'e}les Num{\'e}riques call).}


\begin{thebibliography}{99}

\bibitem{stringari}
L.~P.~Pitaevskii and S.~Stringari,
{\it Bose-Einstein Condensation.}
Oxford University Press (Oxford, 2003).

\bibitem{siambook} P.~G.~Kevrekidis,
D.~J.~Frantzeskakis, and R.~Carretero-Gonz\'alez,
{\it The Defocusing Nonlinear Schr{\"o}dinger Equation},
SIAM (Philadelphia, 2015).


\bibitem{christo} D.~N.~Christodoulides,
Phys. Lett. A, {\bf 132}, 451--452 (1988).

\bibitem{vdbysk1}
V.~V.~Afanasyev, Yu.~S.~Kivshar, V.~V.~Konotop, and V.~N.~Serkin,
Opt. Lett., {\bf 14}, 805--807 (1989).

\bibitem{vddyuri}
Yu.~S.~Kivshar and S.~K.~Turitsyn,
Opt.\ Lett., {\bf 18}, 337--339 (1993).

\bibitem{ralak}
R.~Radhakrishnan and M.~Lakshmanan,
J.\ Phys.\ A: Math.\ Gen., {\bf 28}, 2683--2692 (1995).

\bibitem{dbysk2}
A.~V.~Buryak, Yu.~S.~Kivshar, and D.~F.~Parker,
Phys. Lett. A, {\bf 215}, 57--62 (1996).

\bibitem{shepkiv}
A.~P.~Sheppard and Yu.~S.~Kivshar,
Phys.\ Rev.\ E, {\bf 55}, 4773--4782 (1997).

\bibitem{parkshin}
Q.-H.~Park and H.~J.~Shin,
Phys.\ Rev.\ E, {\bf 61}, 3093--3106 (2000).

\bibitem{seg1} Z. Chen, M. Segev,
T.~H.~Coskun, D.~N.~Christodoulides, and Yu.~S.~Kivshar,
J. Opt. Soc. Am. B, {\bf 14}, 3066--3077 (1997).

\bibitem{seg2}
E.~A.~Ostrovskaya, Yu.~S.~Kivshar, Z.~Chen, and M.~Segev,
Opt.\ Lett., {\bf 24}, 327--329 (1999).


\bibitem{buschanglin}
Th.~Busch and J.~R.~Anglin,
Phys.\ Rev.\ Lett., {\bf 87}, 010401 (2001).

\bibitem{hamburg}
C.~Becker, S.~Stellmer, P.~Soltan-Panahi, S.~D{\"o}rscher, %
M.~Baumert, E.-M.~Richter, J.~Kronj\"{a}ger, K.~Bongs, and %
K.~Sengstock, Nature Phys., {\bf 4}, 496--501 (2008).

\bibitem{pe1}
C.~Hamner, J.~J.~Chang, P.~Engels, and M.~A.~Hoefer,
Phys.\ Rev.\ Lett., {\bf 106}, 065302 (2011).

\bibitem{pe2}
S.~Middelkamp, J.~J.~Chang, C.~Hamner, R.~Carretero-Gonz{\'a}lez, %
P.~G.~Kevrekidis, V.~Achilleos, D.~J.~Frantzeskakis, P.~Schmelcher,%
and P.~Engels, Phys.\ Lett.\ A, {\bf 375}, 642--646 (2011).

\bibitem{pe3}
D.~Yan, J.~J.~Chang, C.~Hamner, P.~G.~Kevrekidis, P.~Engels, V.~Achilleos,
D.~J.~Frantzeskakis, R.~Carretero-Gonz{\'a}lez, and P.~Schmelcher,
Phys.\ Rev.\ A, {\bf 84}, 053630 (2011).

\bibitem{pe4}
M.~A.~Hoefer, J.~J.~Chang, C.~Hamner, and P.~Engels,
Phys.\ Rev.\ A, {\bf 84}, 041605(R) (2011).

\bibitem{pe5}
D.~Yan, J.~J.~Chang, C.~Hamner, M.~Hoefer, P.~G.~Kevrekidis, %
P.~Engels, V.~Achilleos, D.~J.~Frantzeskakis, and J.~Cuevas, %
J.\ Phys.\ B: At.\ Mol.\ Opt.\ Phys., {\bf 45}, 115301 (2012).


\bibitem{azu}
A.~{\'A}lvarez, J.~Cuevas, F.~R.~Romero, C.~Hamner, J.~J.~Chang, %
P.~Engels, P.~G.~Kevrekidis, and D.~J.~Frantzeskakis,
J. Phys. B, {\bf 46}, 065302 (2013).

\bibitem{stringa} C. Qu, L.P. Pitaevskii and S. Stringari,
Phys. Rev. Lett. {\bf 116}, 160402 (2016).

\bibitem{fetter} A.L. Fetter, Phys. Rev. A {\bf 89}, 023629
(2014); see in particular Eq.~(17) for the relevant ansatz.

\bibitem{epjd} P.G. Kevrekidis, H.E. Nistazakis,
D.J. Frantzeskakis, B.A. Malomed and R. Carretero-Gonz{\'a}lez,
Eur. Phys. J. D {\bf 28}, 181 (2004).

\bibitem{kody} K.~J.~H.~Law, P.~G.~Kevrekidis, and L.~S.~Tuckerman,
Phys.\ Rev.\ Lett., {\bf 105}, 160405 (2010).


\bibitem{pola} M.~Pola, J.~Stockhofe,  P.~Schmelcher, and P.~G.~Kevrekidis,
Phys.\ Rev.\ A, {\bf 86}, 053601 (2012).



\bibitem{ringjan} J. Stockhofe, P.G. Kevrekidis, D.J. Frantzeskakis, P. Schmelcher, J. Phys. B At. Mol. Opt. Phys. {\bf 44}, 191003 (2011).

\bibitem{timm} E. Timmermans, Phys. Rev. Lett. 81, 5718 (1998);


\bibitem{Pubig} H. Pu and N. P. Bigelow, Phys. Rev. Lett.
{\bf 80}, 1130 (1998)]




\bibitem{aochui} P. Ao and S. T. Chui
Phys. Rev. A {\bf 58}, 4836 (1998).

\bibitem{rafael} R. Navarro, R. Carretero-Gonz{\'a}lez, and P. G. Kevrekidis
Phys. Rev. A {\bf 80}, 023613 (2009).


\bibitem{kokk} Servaas Kokkelmans, private communication; B. J. Verhaar, 
 E.G.M. van Kempen, and S.J.J.M.F. Kokkelmans,
Phys. Rev. A {\bf 79}, 032711 (2009).



\bibitem{busch2000}
Th.~Busch and J.~R.~Anglin,
Phys.\ Rev.\ Lett., {\bf 84}, 2298 (2000).

\bibitem{konotop}
V.~V.~Konotop and L.~Pitaevskii, Phys.\ Rev.\ Lett., {\bf 93}, 240403 (2004).

\bibitem{hecht-2012-JNM}
F.~Hecht,  Journal of Numerical Mathematics {\bf 20}, 251
(2012).

\bibitem{dan-2010-JCP}
I.~{Danaila}, F.~Hecht, J. Comput. Physics {\bf 229}, 6946 (2010).


\bibitem{stringa3} S. Stringari, Phys. Rev. Lett. {\bf 77}, 2360 (1996).

\bibitem{pelirecent} P.G. Kevrekidis and D.E. Pelinovsky,
Phys. Rev. A {\bf 81}, 023627 (2010).

\bibitem{stephan} S. Middelkamp, P.G. Kevrekidis, D.J. Frantzeskakis,
R. Carretero-Gonz{\'a}lez, and P. Schmelcher,
Phys. Rev. A {\bf 82}, 013646 (2010).

\bibitem{george} G. Theocharis, D. J. Frantzeskakis, P. G. Kevrekidis, B. A. Malomed, and Yu.S. Kivshar
Phys. Rev. Lett. {\bf 90}, 120403 (2003)

\bibitem{kivsharyang} Yu.S. Kivshar and X. Yang, \newblock Phys. Rev. E {\bf 50},
R40 (1994).

\bibitem{dreischuh} D. Neshev, A. Dreischuh, V. Kamenov, I. Stefanov, S. Dinev, W. Fliesser, and L. Windholz,
Appl. Phys. B {\bf 64}, 429 (1997);
A. Dreischuh, D. Neshev, G. G. Paulus, F. Grasbon, and H. Walther,
Phys. Rev. E {\bf 66}, 066611 (2002).

\bibitem{carr} G. Herring, L.D. Carr, R. Carretero-Gonz{\'a}lez, P.G. Kevrekidis, D.J. Frantzeskakis,
Phys. Rev. A {\bf 77}, 023625 (2008).

\bibitem{cluster} S. Middelkamp, P.G. Kevrekidis, D.J. Frantzeskakis, R. Carretero-Gonz{\'a}lez, P. Schmelcher,
Physica D {\bf 240}, 1449 (2011).

\bibitem{stathis} E. G. Charalampidis, W. Wang, P. G. Kevrekidis, D. J. Frantzeskakis, J. Cuevas-Maraver,
 arXiv:1604.04690.




\end{thebibliography}
\end{document}